\documentclass[sigconf,natbib=false]{acmart}
\usepackage{amsmath}

\makeatletter
\def\@ACM@checkaffil{
    \if@ACM@instpresent\else
    \ClassWarningNoLine{\@classname}{No institution present for an affiliation}%
    \fi
    \if@ACM@citypresent\else
    \ClassWarningNoLine{\@classname}{No city present for an affiliation}%
    \fi
    \if@ACM@countrypresent\else
        \ClassWarningNoLine{\@classname}{No country present for an affiliation}%
    \fi
}
\makeatother

\AtBeginDocument{%
  }

\setcopyright{acmcopyright}
\copyrightyear{2023}
\acmYear{2023}
\acmDOI{XXXXXXX.XXXXXXX}

\RequirePackage[
  datamodel=acmdatamodel,
  style=acmnumeric,
  ]{biblatex}

\begin{document}

\title{A Short Note on Setting Swap Parameters}

\author{Nihar Shah}
\email{nshah@jumpcrypto.com}
\affiliation{%
  \institution{Jump Crypto}
}

\author{Lucas Baker}
\email{lbaker@jumpcrypto.com}
\affiliation{%
  \institution{Jump Crypto}
}

\author{Suraj Srinivasan}
\email{ssrinivasan@jumpcrypto.com}
\affiliation{%
  \institution{Jump Crypto}
}

\author{Alex Toberoff}
\email{atoberoff@jumpcrypto.com}
\affiliation{%
  \institution{Jump Crypto}
}

\renewcommand{\shortauthors}{Shah, Baker, Srinivasan, and Toberoff}

\begin{abstract}
  This short note illustrates the theoretical solution to a trader determining how to optimally swap her wealth into a target asset through on-chain operations. It offers the framework to solve optimal slippage parameters and optimal trade size.
\end{abstract}



\settopmatter{printfolios=true}
\maketitle

\section{Introduction}
How should on-chain participants set parameters when making swaps on decentralized trading platforms? Should they set slippage tolerances low or high? Should they make large trades or split them into small batches? In too many cases, this is done in an ad-hoc manner with rules of thumb, causing transactions to either fail or leak value to MEV bots.

This is a complicated question, but this short note makes some progress. In particular, it provides a theoretical solution for one interesting, if stylized case, and illustrates how a high-dimensional problem can be solved and represented succinctly.

This does not immediately translate into practical outcomes. Indeed, further empirical analysis would be needed to calibrate, test, and operationalize the solution. However, protocol designers and practitioners alike can continue building on this corpus of work to answer the opening question in a robust manner.

\section{General Setting}
The agent in our model is a trader who wants to swap some endowed wealth into an alternate asset. For simplicity, the endowed wealth $W$ is the native asset of the chain, while the target asset has some price $p$. The agent faces a single liquidity pool but can swap this position over as many blocks as she likes. She pays a constant gas cost $g$ to launch a trade, regardless of whether it is successful.

The trader's objective function can be formalized below. In words, she operates over $T$ time steps but begins formulating her strategy at $t = 0$. At each time step, she accrues a position in the target asset $x_t$. At the conclusion of the game (time $T$), she assesses the total notional sum of those positions and thus multiplies the summation by the price at that point $p_T$ (as she cares about the terminal value of her position).
$$
\max \mathbb{E}_0\left[p_T \left(\sum_{t = 1}^{T} \delta^t x_t\right)\right]~~~\text{where}~\lim T \rightarrow \infty~~\text{and}~~\lim \delta \rightarrow 1
$$

This objective function embeds two minor mathematical adjustments. First, while the game has a finite horizon, we approximate it with infinite time. Second, we embed an infinitesimally small discount rate inside the objective function. The first adjustment allows us to ignore terminal conditions while ensuring that the terminal price remains a well-defined concept. The second adjustment prevents the trader from getting stuck in a world with multiple equivalent equilibria (i.e., between trading today and trading tomorrow) and allows us to ensure the trader starts executing her optimal strategy immediately.

We deliberately omit introducing the budget constraint until the next section. To solve this problem, we must solve a closely-related sub-problem.

\section{One-Shot Setting}
The sub-problem of the one-shot setting inherits from the general setting, but it embeds three simplifications. First, we assume that the trader wishes to swap some wealth $w$ in a \textit{single} trade. Second, we assume that future liquidity conditions and prices cannot be forecasted, and so the trader always decides to execute that single trade immediately. Third, we assume that the gas cost $g$ the trader pays to execute a trade is paid out of a separate account with no funding constraints. Thus, even if the trader's trades fail, the wealth she attempts to swap remains constant (i.e., she continues swapping $w$ rather than $w - g$ if the first attempted trade fails). This raises the possibility of a negative value in the objective function, but this does not change the solution process.

This yields the following objective function, where $I(\cdot)$ refers to a random indicator variable. For brevity, we omit the asymptotic conditions around $T$ and $\delta$.
$$
\max \mathbb{E}_0\left[p_T \left(\sum_{t = 1}^{T} \delta^t x_t\right) - \sum_{t = 1}^{T} g I(x_t > 0)\right]
$$

Moreover, because the trader swaps her full wealth $w$ into $x$ starting immediately, we can define $x_t$ in terms of her budget. We demonstrate two such budget constraints, corresponding to $x_1$ and $x_2$, below. The budget constraints embed four new terms, which are explained next.
$$
x_1 = I(Z(\Omega_1, w) < c_1) \frac{w}{M(\Omega_1, w, c_1)}
$$
$$
x_2 = I(Z(\Omega_1, w) > c_1) \left(I(Z(\Omega_2, w) < c_2) \frac{w}{M(\Omega_2, w, c_2)}\right)
$$

The first variable to introduce is the cutoff price $c_t$, the worst-case price the trader chooses to get filled for that block. In short, it is the trader's slippage tolerance.

The second variable is $\Omega_t$, which reflects all exogenous parameters that govern the liquidity state of the pool at the point in time of the trade.

The third term is the pricing function $Z(\cdot)$ that reflects the average \textit{exogenous} price that the trader would pay, as a function of the pool's liquidity and the trade size. This assumes no MEV bots try to extract value from the trader's swap.

The fourth term is the pricing function $M(\cdot)$ that reflects the average \textit{endogenous} price that the trader would pay, as a function of the pool's liquidity, the trade size, and the trade cutoff value. In contrast to the previous term, this term explicitly accounts for the possibility of MEV bots extracting value from the trader's swap. Note that it will always be weakly bounded by the slippage-tolerant price $c_t$ if the trade is executed, although it can be lower.

A full analysis of $M(\cdot)$ is out of scope for this paper -- but Heimbach and Wattenhofer (2022) offer a richer treatment.\cite{heimbach} They discuss in more detail the conditions under which front-running by MEV bots (often known as ``sandwich attacks") emerge and push up the price paid by the trader. Some combination of theoretical and empirical analysis should be used to chart this function.

These four terms, used in conjunction with the indicator function $I(\cdot)$, can be used to condition the value of $x_1$ on whether the first trade was successful, i.e., whether the exogenous price was below the cutoff value. We similarly only allow $x_2$ to be potentially non-zero if the first trade was unsuccessful. This yields a general solution for a given position in a given time period $x_t$. This position is always zero unless all previous trades failed and this trade was successful.
$$
x_t = \left(\prod_{1 \le j < t} I(Z(\Omega_j, w) > c_j)\right) \left(I(Z(\Omega_t, w) < c_t) \frac{w}{M(\Omega_t, w, c_t)}\right)
$$

We now write the objective function that directly embeds these constraints, where the decision variable has changed from the position $x_t$ to the slippage threshold $c_t$. The gas component can also be written in a simplified manner: it is zero unless all previous trades have failed. We can omit the infinitesimal discounting since the trader executes her strategy immediately, i.e., attempts to swap her wealth into the target asset.
\begin{multline}
\max_{\{c_t\}} \mathbb{E}_0 \left[\sum_{t = 1}^{T} \left(\prod_{1 \le j < t} I(Z(\Omega_j, w) > c_j) \right) \times \right.  \\
\left. \left( p_T \left( I(Z(\Omega_t, w) < c_t) \frac{w}{M(\Omega_t, w, c_t)} \right) - g \right) \right]
\label{eq:core}
\end{multline}

We simplify the problem in two ways. First, we assume price and liquidity are martingale processes that evolve through a mean-zero vector error $\eta$. They are cointegrated, but the trader has no information on their future trajectory and thus uses the current values as the forecast. While this additive structure may not be ideal for long-run horizons, it seems reasonable at short horizons.
$$
\left( \begin{array}{c}
     p_{t + 1} \\
     \Omega_{t + 1} 
\end{array} \right) = 
\left( \begin{array}{c}
     p_{t} \\
     \Omega_{t} 
\end{array} \right) + \eta_{t + 1}
$$

Second, we give the decision variable space some time-invariant structure, where the price cutoff $c$ is always set to be some constant distance $a$ away from the current price $p$. In other words, the trader takes the current price, adjusts it upwards by some constant, and uses this as the cutoff for the next period. This compresses the trader's decision variable from a set to a single value.
$$
c_{t+1} = p_t + a
$$

These simplifications make the problem more tractable. We rewrite Equation \eqref{eq:core} with these adjustments. We further let $T \rightarrow \infty$, rearrange terms, and employ the law of iterated expectations.
\begin{multline*}
\max_{a} \sum_{t = 1}^{\infty} \mathbb{E}_0 \left[ \mathbb{E}_{t - 1} \left[ \left(\prod_{1 \le j < t} I(Z(\Omega_j, w) > p_{j - 1} + a) \right) \times \right. \right. \\
\left. \left. \left( \left( I(Z(\Omega_t, w) < p_{t - 1} + a) \frac{w \left(p_t + \sum_{i = t + 1}^{\infty} \eta_{i,1} \right)}{M(\Omega_t, w, p_{t - 1} + a)} \right) - g \right) \right] \right]
\end{multline*}

We can then drag the expectation operators through the terms. From the perspective of the $\mathbb{E}_{t - 1}$ operator, all uncertainty on whether previous prices have been higher than previous cutoffs have been resolved. From the perspective of that operator, we can also eliminate future $\eta_{t + 1}$ terms, as they are independent random variables that are mean-zero.
\begin{multline*}
\max_{a} \sum_{t = 1}^{\infty} \mathbb{E}_0 \left[ \left(\prod_{1 \le j < t} I(Z(\Omega_j, w) > p_{j - 1} + a) \right) \times \right. \\
\left. \mathbb{E}_{t - 1} \left[ I(Z(\Omega_t, w) < p_{t - 1} + a) \frac{w p_t}{M(\Omega_t, w, p_{t - 1} + a)} - g \right] \right]
\end{multline*}

We make two simplifications based on the stationarity of the price, liquidity, and cutoff processes. First, we argue that the distribution of $Z_t - p_{t - 1}$ is identical for any future $t$ (for a given swap size $w$). Thus, we define the residual as the random variable $\epsilon_t$, with the associated cumulative density function $F_w(\cdot)$. Second, we argue that the ratio $p_t/M_t$ has the same expected value taken in the previous time step, for any $t$. While this holds unconditionally, we argue that it critically holds conditional on the trade being executed. These steps allow us to simplify the inner term into a conditional expectation term, which can also be removed from the summation as it is constant across all future time periods (and thus equivalent to the expectation taken at $t = 0$).
\begin{multline*}
\max_{a} \left( w F_{w}(a) \mathbb{E} \left(\frac{p}{M(\Omega, w, p_{-1} + a)} \rvert \epsilon < a \right) - g \right) \times \\
\left(\sum_{t = 1}^{\infty} \mathbb{E}_0 \left[ \prod_{1 \le j < t} I(\epsilon_j > a) \right] \right)
\end{multline*}

The expectation on the right-hand side can be converted into a product of cumulative density functions of $\epsilon$.
$$
\max_{a} \left( w F_{w}(a) \mathbb{E} \left(\frac{p}{M(\Omega, w, p_{-1} + a)} \rvert \epsilon < a \right) - g \right) \left(\sum_{t = 1}^{\infty} (1 - F_{w}(a))^{t - 1} \right)
$$

This term on the right-hand side should be recognized as a geometric sequence, by re-indexing the sequence to start at $t = 0$. This simplifies into $F_{w}(a)^{-1}$ and allows for a straightforward simplification of our final result.
\begin{equation}
\max_{a} w \mathbb{E} \left(\frac{p}{M(\Omega, w, p_{-1} + a)}  \rvert \epsilon < a \right) - \frac{g}{F_{w}(a)}
\label{eq:simple}
\end{equation}

Equation \eqref{eq:simple} is the resulting objective function, which a protocol or practitioner can optimize empirically. For a given exogenous price process for the AMM $F_w(\cdot)$ and for a given endogenous pricing function $M(\cdot)$, the optimal slippage tolerance $a$ for a single-slot trade can be computed.

Equation \eqref{eq:simple} is also insightful and simple in its own right. Importantly, it illustrates the core trade-off. An increase in $a$ lowers the expected value in the first term by making the trade more prone to high slippage (either due to limited liquidity or extraction by an MEV bot). But an increase in $a$ also lowers the second term (i.e., raises the denominator), and so fewer gas costs are paid. Indeed, this second term can also be interpreted directly as the expected value of a hypergeometric distribution, whereby gas costs are paid until the trade succeeds (in each iteration, with $F_w(a)$ probability of success).

\section{Returning to the General Setting}
The previous section solves the problem when the trader swaps all wealth $w$ in a single trade. For some practitioners, this is sufficient. However, we now return to the general problem of swapping some total wealth $W$ through a single pool over some number of trades.

The core insight is to continue leveraging the stationarity assumption for that pool and thus split $W$ into a set of equally-sized $w$ to execute sequentially. Again, it is worth stressing that the stationarity assumption will not hold perfectly -- as one successful swap will impair the liquidity state for future swaps -- but this approximation can be considered reasonable if the pool is large and active relative to the trader's position. In practice, traders may wish to make ad-hoc adjustments based on the results of empirical simulations or realized experiences.

One approach is to simply utilize an iterative search. For instance, Equation \eqref{eq:simple} can be checked on whether one trade that takes $W$ or two trades that take $W/2$ yields more value, and this can be repeated iteratively (e.g., three trades that take $W/3$). This is especially useful for small amounts of wealth, where the search space is low-dimensional. However, there is a more general solution.

In particular, suppose we have an approach for optimizing Equation \eqref{eq:simple}. We can use this to construct a value function $V(w)$, which takes in a given level of wealth and returns the value of the objective function, evaluated at the optimal $a$ for that particular parameterization.
\begin{equation}
V(w) = \sup \left[ w \mathbb{E} \left(\frac{p}{M(\Omega, w, p_{-1} + a)} \rvert \epsilon < a \right) - \frac{g}{F_{w}(a)} \right]
\label{eq:value}
\end{equation}

Now, assuming some concavity in $V(\cdot)$ with respect to wealth, we find the point in that function where returns to additional wealth begin to diminish steeply. (This concavity stems from a larger trade size being more lucrative for MEV bots.) In particular, we find the point below -- where 1\% more wealth in a given trade and 1\% fewer trades (and vice-versa) is suboptimal to any given bundle. This is equivalent to taking the derivative of a constrained maximization problem, but this particular presentation may be easier to understand and operationalize.
\begin{equation}
w^*: V(w) \ge kV(w/k)~\text{as}~k \rightarrow 1^{+}~\text{and}~k \rightarrow 1^{-}
\label{eq:condition}
\end{equation}

The result of Equation \eqref{eq:condition} gives us the optimal trade size. Interestingly, this does not need to be computed in a bespoke manner for a given total wealth level $W$. As long as that wealth is ``large enough" that any fractional leftover wealth is negligible, each trade size should be $w^*$. In practice, fractional wealth will surely exist because failed trades will lead to extra gas consumed; but we treat that remainder as negligible to the problem.

This concludes the exercise. Under some parameters, some functional forms, and some assumptions, we can compute $w^*$ and recommend all traders launch each trade as no larger or smaller than that. They use this solution and repeat until they swap all their wealth into the target asset, and this strategy is \textit{regardless} of their initial wealth.

\section{Extensions}
There are many extensions of the framework. We outline three.

First, this model assumes that total wealth is sufficiently large. But if total wealth is small, there are two potential adjustments. First, the iterative search outlined in the previous section may be preferable. The trader should consider which of the following options yields the highest total value: a single trade with total wealth, two trades with half wealth, three trades with a third wealth, and so on. Second, the trader may wish to ensure that her total ending wealth is not negative in any state of the world, and thus, she sets more permissive slippage parameters as her wealth stock dwindles. Put another way, the trader may be locally risk-averse around zero wealth, necessitating a change in the objective function and optimal strategy.

Second, this model assumes a trader wishes to swap through a single pool over multiple blocks. There is a related question: a trader wishes to swap through multiple pools across a single block. As long as wealth $W$ is sufficiently large, the solution can proceed similarly. In particular, we can specify multiple $V(\cdot)$ functions. (As an example, we specify a $V_1(\cdot)$ function and an optimized $w_1^*$, a $V_2(\cdot)$ function and optimized $w_2^*$, and so on.) For enough pools, this yields an optimal set that is irrespective of total wealth. The trader then swaps across multiple pools in tandem, until all wealth is exhausted. This approach can get complicated if trades are risk-averse (as failures across multiple pools are correlated), if total wealth is small (requiring some solution adjusted through a grid search), or if the stationarity assumption is particularly broken (as taking liquidity from multiple pools concurrently may not be immediately replenished).

Third, this model assumes total wealth is constant rather than stochastic. That being said, if wealth is stochastic, the model does not meaningfully change. The key insight is that the optimal trade size is fixed regardless of initial wealth, and so fluctuations in the total wealth $W$ do not change that insight.

\section{Conclusion}
This note illustrates a solution to the problem of setting slippage parameters by compressing a complex problem into a simple one. While the setting is stylized, the note shows how some simple assumptions can make an intractable problem tractable.

Future work by protocols and practitioners can build upon this. While any practical solution will require empirical work, this note provides the framework for guiding the analysis.


\printbibliography



\end{document}